\documentclass[preprint,onecolumn,amsmath,showkeys]{revtex4}

\usepackage{graphicx}
\usepackage{dcolumn}
\usepackage{bm}
\usepackage[T1]{fontenc}

\begin{document}

\title{Molecular Dynamics Study of Ferroelectric Perovskites based on Effective Hamiltonians: implementation of Nos\'e-Hoover and Parinello-Rahman algorithms.}

\author{Gr\'egory Geneste\footnote{electronic address : gregory.geneste@ecp.fr}}
\address{Laboratoire Structures, Propri\'et\'es et Mod\'elisation des Solides, CNRS-UMR 8580,
Ecole Centrale Paris, Grande Voie des Vignes, 92295 Ch\^atenay-Malabry Cedex, France\\}

\begin{abstract}
Molecular Dynamics is applied to Ferroelectric Perovskites in the framework of a first-principles derived effective Hamiltonian (Zhong, Vanderbilt, Rabe, Phys. Rev. Lett. {\bf 73} (1994), 1861). The degrees of freedom, that obey the Newton equations of motion, are the local modes and the displacement modes. The Nos\'e-Hoover method is implemented, as well as the Parinello-Rahman scheme to perform fixed temperature and fixed stress tensor simulations. This allows to study the thermodynamics of ferroelectric perovskites and to reproduce successfully the Monte Carlo results on phase transitions, polarization and homogeneous strain evolution with temperature of BaTiO$_3$, taken as an example. 
\end{abstract}

\keywords{Molecular Dynamics, Effective Hamiltonian, Nos\'e-Hoover, Parinello-Rahman}

\maketitle

\section{Introduction}
\label{intro}
The study of ferroelectric (FE) and related perovskite oxides has been the subject of many atomic-scale theoretical and numerical studies for at least fifteen years. In the ninetees, there has been an increasing growth of numerical work due to the appearance of significant computational facilities. Ab initio calculations have considerably developed, and have been used very early to derive parameters of phenomenological models used to understand and predict the thermodynamics of phase transitions in perovskite oxides, that can be quite complex.

One of the most powerful tool, that captures most of the thermodynamics of FE materials, is the so-called "Effective Hamiltonian" (EH), introduced by Zhong, Vanderbilt and Rabe\cite{zhong94,zhong95} in 1994.
It can be viewed as a simplification of the potential energy surface of a ferroelectric in terms of relevant degrees of freedom such as the local modes, the inhomogeneous strain tensor and the homogeneous strain tensor. The expression of this hamiltonian is phenomenological and all the parameters are directly derived from first-principles DFT calculations, usually performed in the framework of the Local Density Approximation (LDA). Given this hamiltonian, one can perform Monte Carlo (MC) simulations to predict, at given temperature and pressure (or stress tensor), various physical quantities obtained as statistical averages (polarization, strain, dielectric constant, etc). Zhong, Vanderbilt and Rabe used it to reproduce successfully the complex sequence of phase transitions of barium titanate\cite{zhong94,zhong95}.

The MC framework is a numerical technique that allows to sample the space of configurations according to an equilibrium probability distribution. An alternative technique, that normally yields similar results, is the Molecular Dynamics (MD). In MD, the degrees of freedom evolve with time according to the Newton equations of motion, leading, if the trajectory is long enough, to an equilibrium sampling of the phase space. Thus, time-averaged microscopic quantities on such equilibrium trajectories can be used to obtain macroscopic quantities, that should be the same as those obtained within MC simulations (ergodicity hypothesis).

MD has already been used with the Effective Hamiltonian. For instance, Burton {\it et al}\cite{burton_md} applied it to the simulation of relaxor materials. Recently, Ponomareva {\it et al} applied it to the study of THz dielectric response of barium titanate\cite{ponomareva08}.

In the present work, we apply the MD method to the Effective Hamiltonian. We implement the Nos\'e-Hoover and Parinello-Rahman schemes, that allow to perform MD with fixed temperature and fixed stress tensor. We first recall in the theoretical background the basis of the Effective Hamiltonian introduced by Zhong, Vanderbilt and Rabe, and describe the Nos\'e-Hoover and Parinello-Rahman algorithms. Then the method is successfully applied to barium titanate to reproduce results very similar to those obtained by MC simulations.

\section{Theoretical background}
\subsection{Degrees of Freedom}
The main idea of the effective hamiltonian is to decrease the number of degrees of freedom (a priori 15 per ABO$_3$ unit cell in a perovskite) by using new coordinates representing collective motions of the atoms in each unit cell $i$.

We assume that the considered perovskite oxides have an energy landscape which is well described in the framework of the effective hamiltonian proposed by Zhong {\it et al} in 1994\cite{zhong94,zhong95}. This hamiltonian deals with three kinds of degrees of freedom:

(i) the local modes $\vec u_i$, that roughly represent the polar displacements of the atoms of unit cell $i$, which induce an electric dipole in each unit cell. The dipole is obtained as $Z^* \vec u_i$, where $Z^*$ is the effective charge of the soft mode, calculated in the cubic structure.

(ii) the inhomogeneous displacement modes $\vec v_i$, that describe the local strain in cell $i$ and refer to long-wavelength acoustic modes. From this displacement field, one constructs easily the inhomogeneous strain tensor field $\eta_l^I (i)$\cite{zhong95}.

(iii) the homogeneous strain tensor $\eta_l^H$, that does not depend on the unit cell and is a global deformation applied to the whole system.

The strain existing at cell $i$ is therefore $\eta_l(i) = \eta_l^H + \eta_l^I (i)$. 

In many perovskite oxides, other atomic motions arising from the freezing of zone-boundary modes play a very important role. These acoustic modes at the $R$ or $M$ points of the First Brillouin Zone of the cubic structure, called antiferrodistortions (AFD), consist of rotations of the oxygen octahedra around a given axis. These modes are not considered here but could be dealt with exactly in the same way as local modes and displacement modes.
In the following, we assume that the system studied consists of N unit cells. Thus the number of internal degrees of freedom is 6N.

\subsection{Effective Hamiltonian}

The form proposed by Zhong {\it et al} in 1994\cite{zhong94,zhong95} is :

\begin{eqnarray*}
&& \mathrm H^{eff}(\left\{\vec u_i \right\},\left\{\eta_l^I(i)\right\},\left\{\eta_l^H \right\})  = E^{self}(\left\{\vec u_i \right\}) \\
&& +  E^{dpl}(\left\{\vec u_i \right\}) +  E^{short}(\left\{\vec u_i \right\})  \\
&& + E^{elas}(\left\{\eta_l^I(i)\right\}, \left\{\eta_l^H \right\}) + E^{int}(\left\{\vec u_i \right\}, \left\{\eta_l^I(i)\right\}, \left\{\eta_l^H \right\}),
\end{eqnarray*}

in which we can find a local mode self-energy $E^{self}$ (acting as a local potential for the local modes), a dipole-dipole electrostatic interaction energy $E^{dpl}$, a short-range interaction energy $E^{short}$ (extended up to third-neighbors), an elastic energy $E^{elas}$ (depending only on the homogeneous and inhomogeneous strains) and a term coupling the strain to the local modes $E^{int}$ (crucial to describe piezoelectric effects). The precise expression of all these terms is given in Ref.~\onlinecite{zhong95}.

Interestingly, this hamiltonian only includes local anharmonicities in the onsite part and in the coupling between local modes and strain.

\subsection{Dynamics}
We assume that a mass can be affected to the local modes ($m_{lm}$) and displacement modes ($m_{dsp}$). It is thus possible to formulate time-evolution equations for these quantities:

\begin{equation}
m_{lm}\frac{d^2 \vec u_i}{dt^2} = - \frac{\partial H^{eff}}{\partial \vec u_i} = \vec f_i^{lm}
\end{equation}

\begin{equation}
m_{dsp}\frac{d^2 \vec v_i}{dt^2} = - \frac{\partial H^{eff}}{\partial \vec v_i} = \vec f_i^{dsp}
\end{equation}

where $lm$ and $dsp$ stand respectively for local mode and displacement mode. In the code we have implemented, the forces are calculated directly according to an analytical derivation of the effective hamiltonian (taken exactly as it is in Ref.~\onlinecite{zhong95}).
The masses $m_{lm}$ and $m_{dsp}$ are chosen according to Ref.~\onlinecite{nishimatsu2008}.

\section{Molecular Dynamics}
The previous equations of motion are numerically integrated within the well-known Verlet algorithm\cite{verlet}. A time step $h$ is chosen, usually in the range $10^{-15}$s. In this scheme, the local mode at time $t+h$ is computed from the local modes at $t$ and $t-h$, using the forces calculated at $t$:

\begin{equation}
\vec u_i(t+h) = 2 \vec u_i(t) - \vec u_i(t-h) + \frac{h^2}{m_{lm}} \vec f_i^{lm}(t)
\end{equation}

and idem for the displacement modes:
 
\begin{equation}
\vec v_i(t+h) = 2 \vec v_i(t) - \vec v_i(t-h) + \frac{h^2}{m_{dsp}} \vec f_i^{dsp}(t)
\end{equation}

Of course, at each step, the different degrees of freedom are coupled and are coupled to the homogeneous strain:

\begin{equation}
\nonumber
\vec f_i^{lm}(t)=\vec f_i^{lm}(\left\{\vec u_j(t) \right\},\left\{\eta_l^I(j)(t)\right\},\left\{\eta_l^H (t) \right\})
\end{equation}

The dynamics is started from initial positions and from initial velocities, chosen randomly to reproduce the Maxwell distribution corresponding to an initial temperature, as usual in MD.

For the first step, the positions and velocities are computed from a simple Taylor expansion:
\begin{equation}
\nonumber
\vec u_i(h) = \vec u_i(0) + h \frac{d \vec u_i}{dt}(0) + \frac{h^2}{2 m_{lm}} \vec f_i^{lm}(0)
\end{equation}
\begin{equation}
\nonumber
\vec v_i(h) = \vec v_i(0) + h \frac{d \vec v_i}{dt}(0) + \frac{h^2}{2 m_{dsp}} \vec f_i^{dsp}(0)
\end{equation}

The instantaneous temperature is calculated from the velocities, based on equipartition theorem. 
\begin{eqnarray*}
\nonumber
&& \mathrm 2 \times \frac{3}{2} N k_B T(t) = \sum_i \frac{1}{2} m_{lm} \left\{ \frac{d \vec u_i}{dt}(t)\right\}^2  \\
&& +  \sum_i \frac{1}{2} m_{dsp} \left\{ \frac{d \vec u_i}{dt}(t)\right\}^2
\end{eqnarray*}

The computation of the electrostatic dipole-dipole energy and electrostatic dipole-dipole forces is performed by the Ewald method, following the scheme proposed by Zhong {\it et al}\cite{zhong95}.

\subsection{Microcanonic molecular dynamics (NVE)}
In the microcanonic scheme, the homogeneous strain is fixed and does not vary all along the simulation:
$\eta_l^H(t)=\eta_l^H$. The temperature can not be chosen a priori and is found at the end of the simulation as the time average of the "instantaneous temperature".

\subsection{Nos\'e-Hoover molecular dynamics (NVT)}
The Nos\'e-Hoover algorithm is a method to fix the equilibrium temperature to the desired temperature T$_0$\cite{nose84,nose86,hoover85}. We describe briefly this method in this section.
A fictitious degree of freedom $s$ with "mass" $Q$ is added that represents a thermostat. The equations of motion are modified according to:

\begin{equation}
\nonumber
m_{lm}\frac{d^2 \vec u_i}{dt^2} =  \vec f_i^{lm} - m_{lm} \zeta (t) \frac{d \vec u_i}{dt}
\end{equation}

\begin{equation}
\nonumber
m_{dsp}\frac{d^2 \vec v_i}{dt^2} =  \vec f_i^{dsp} - m_{dsp} \zeta (t) \frac{d \vec v_i}{dt}
\end{equation}

\begin{equation}
\nonumber
\frac{d \zeta}{dt} =  - \frac{k_B g T(t)}{Q} (\frac{T_0}{T(t)} -1)
\end{equation}

where T(t) and T$_0$ are respectively the instantaneous and targetted temperature, $g$ is the number of degrees of freedom in the system, and $\zeta (t)= \frac{d ln s}{dt} (t)$. $Q$ has not the dimension of a mass, but is the product of an energy by the square of a time. Its value must be appropriate so that the exchange of energy between the system and the thermostat are neither too slow nor too fast. Anyway, when chosen correctly, the precise value of $Q$ does not influence the quality of the phase space sampling and has no effect on the macroscopic averages performed on an equilibrium trajectory.

These equations are also numerically integrated within the Verlet algorithm:

\begin{equation}
\nonumber
\vec u_i(t+h) = 2 \vec u_i(t) - \vec u_i(t-h) + \frac{h^2}{m_{lm}} (\vec f_i^{lm}(t) - m_{lm} \zeta(t) \frac{d \vec u_i}{dt}(t) )
\end{equation}

in which the velocity at time step $t$ is estimated according to\cite{ferr85}:

\begin{equation}
\label{vel-ferr85}
\frac{d \vec u_i}{dt}(t) = \frac{3 \vec u_i(t) -4 \vec u_i(t-h) + \vec u_i(t-2h)}{2 h}
\end{equation}

\subsection{Parinello-Rahman molecular dynamics}
The Parinello-Rahman algorithm\cite{parinello} is a very efficient method to perform molecular dynamics under fixed stress tensor. Such a method is crucial to study phase transitions in FE oxides since various phases currently appear (cubic, tetragonal, orthorhombic, rhomboedral, etc). 
In the Parinello-Rahman scheme, the supercell is allowed to vary in shape and volume along the simulation so that the macroscopic stress tensor (average of the instantaneous stress tensor) is equal to the desired one.
We summarize the Parinello-Rahman method in the following, as we have implemented it.

We denote by $H(t)$ the $3 \times 3$ matrix formed by the components of the three vectors $\vec a$, $\vec b$ and $\vec c$ that define the supercell of the simulation, expressed in an absolute orthonormal coordinate system, attached to the cubic 5-atom unit cell. It depends on $t$ since the supercell is allowed to evolve in shape and volume along the simulation.
$H(t)$ is related to the homogeneous strain tensor $\eta^H(t)$.

The matrix $G(t)$ is defined by $G(t) = ^tH(t).H(t)$.
The local mode at $(i)$ has three components $u'_{i \alpha}$ ($\alpha$=x,y,z) in the time-dependent basis ($\vec a$, $\vec b$, $\vec c$) and we denote by $\vec u'_i$ the vector with components $u'_{i \alpha}$.
We have $\vec u_i = H(t).\vec u'_i$, where $\vec u_i$ is the vector formed by the components of the local mode at $(i)$ in the absolute coordinate system.
In the Parinello-Rahman framework, the equations of motion are formulated on the scaled variable $u'_{i \alpha}$:

\begin{equation}
\nonumber
m_{lm}\frac{d^2 \vec u'_i}{dt^2} =  H(t)^{-1}.\vec f_i^{lm} - m_{lm} G(t)^{-1}.\frac{dG}{dt}(t) \frac{d \vec u'_i}{dt}
\end{equation}

With the same notations for the displacement modes $\vec v_i = H(t).\vec v'_i$,

\begin{equation}
\nonumber
m_{dsp}\frac{d^2 \vec v'_i}{dt^2} =  H(t)^{-1}.\vec f_i^{dsp} - m_{dsp} G(t)^{-1}.\frac{dG}{dt}(t) \frac{d \vec v'_i}{dt}
\end{equation}

The $H(t)$ matrix evolves according to:

\begin{equation}
\nonumber
W \frac{d^2 H}{dt^2}(t) = (\underline{\sigma} - \underline{\sigma_{0}}).\underline{\omega}
\end{equation}

in which $W$ is a "mass" associated to the dynamics of the unit cell vectors $\vec a$, $\vec b$ and $\vec c$, $\underline{\omega}_{ij} = \partial \Omega / \partial H_{ij}$ with $\Omega$ the volume of the MD supercell, $\underline{\sigma}$ is the instantaneous stress tensor and $\underline{\sigma_{0}}$ the targetted stress tensor. The mass $W$ has to be chosen correctly, in the same manner as $Q$. Its precise value, when chosen in the correct range, does not influence the thermal averages over equilibrium trajectories\cite{parinello}.

Here again the equations of motion are integrated within the Verlet algorithm. The velocities are calculated at each step according to Eq.~\ref{vel-ferr85}.
Using this algorithm, the instantaneous stress tensor components $\underline{\sigma}_{ij}$ oscillate around the targetted value $\underline{\sigma_0}_{ij}$ so that their average over an equilibrium trajectory is equal to $\underline{\sigma_0}_{ij}$.

The instantaneous stress tensor is calculated in this work by
\begin{equation}
\label{stress-tensor}
\underline{\sigma}_{ij} = \frac{1}{\Omega} \frac{\partial H^{eff}}{\partial \epsilon_{ij}},
\end{equation}

where $\Omega$ is the supercell volume and $\epsilon_{ij}$ the $(i,j)$-component of the homogeneous strain tensor.

\section{Example: BaTiO$_3$}
A code has been written in fortran 90, that includes microcanonic, Nos\'e-Hoover, Parinello-Rahman and also both Nos\'e-Hoover and Parinello-Rahman algorithms, that can be coupled to perform fixed temperature and fixed stress tensor MD\cite{hoover85}. This is precisely what is required to study the thermodynamics of ferroelectric materials and perform thermal averages. We present here the results obtained on the prototypical example of BaTiO$_3$ (BTO).
The following computations have been performed with a time step of 2$\times$10$^{-15}$s on a supercell 12 $\times$ 12 $\times$ 12 with periodic boundary conditions.

Barium titanate, the prototypical ferroelectric oxide, has been the first solid to which the formalism of effective hamiltonian was applied in 1994, together with Monte Carlo simulations. We use here the numerical parameters given in Ref.~\onlinecite{zhong95}. The local modes are initialized to zero, as well as the displacement modes and the homogeneous strain. Each equilibrium trajectory is obtained independently from the others ({\it i.e.} the simulations are not initialized with the result of a previous computation).

The simulations are performed in the Nose-Hoover and Parinello-Rahman scheme with an imposed diagonal stress tensor corresponding to a negative hydrostatic pressure P = -4.8 GPa ({\it i.e.} positive stress tensor components). This negative pressure is the one used by Zhong {\it et al} in their simulations to correct the underestimation of the lattice constant by the LDA.
We perform simulations of 10$^5$ steps. During the 50 000 first steps, the system is equilibrating, and the averages are computed on the 50 000 last steps.
In fact, the thermodynamic equilibrium is in general quickly reached (within a few thousands steps).

\begin{figure}[htbp]
    {\par\centering
    {\scalebox{0.30}{\includegraphics{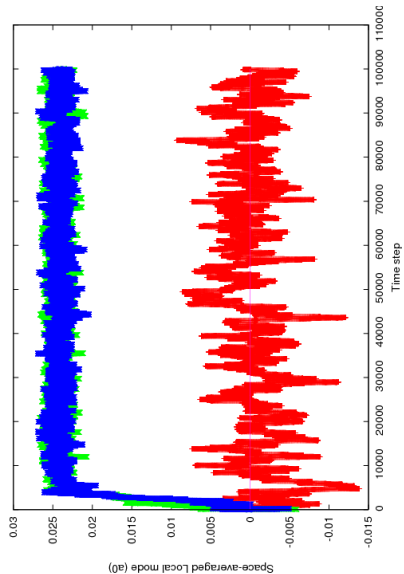}}}
    \par}
     \caption{{\small Time evolution of the three components of the local modes (averaged over the whole simulation cell) at T=210K.}}
    \label{polar200}
\end{figure}

We first give an example at T=210K, for which BTO is found orthorhombic. We examine the time evolution of the space-averaged homogeneous strain tensor components and of the space-averaged local modes (proportionnal to the instantaneous polarization). After $\approx$ 4000 steps, two components of the average local mode dissociate from the third and reach the value of $\approx$ 0.024 a$_0$, while the third one oscillates around zero. The 10$^5$ steps of this evolution are shown on Fig.~\ref{polar200}. Note that two states of macroscopic polarization (positive and negative) are of course possible for each component and degenerate.

It can be noted that the fluctuations of the vanishing component are, as expected, much larger than the two others.

\begin{figure}[htbp]
    {\par\centering
    {\scalebox{0.30}{\includegraphics{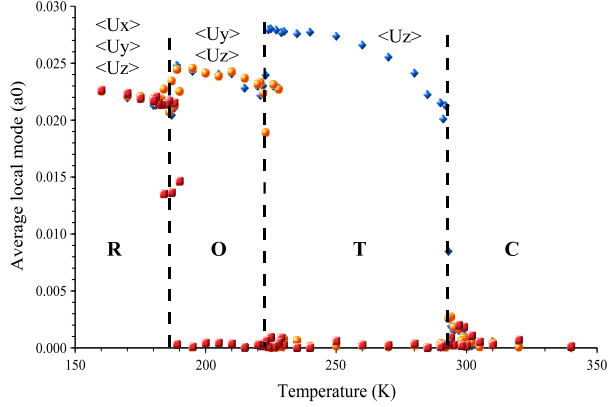}}}
    \par}
     \caption{{\small Average local mode components as a function of temperature. R, O, T and C denote respectively the Rhombohedral, Orthorhombic, Tetragonal and Cubic phases of BaTiO$_3$. Around phase transitions, on a small temperature range $\approx$ 10 K, either one phase or the other can be obtained since all the points are obtained independently from each other (The MD is in each case initialized in the same way).}}
    \label{polarization}
\end{figure}

\begin{figure}[htbp]
    {\par\centering
    {\scalebox{0.30}{\includegraphics{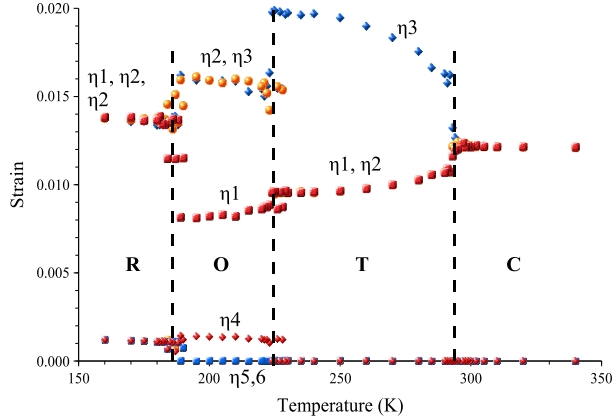}}}
    \par}
     \caption{{\small Average strain tensor components as a function of temperature (Voigt notation). R, O, T and C denote respectively the Rhombohedral, Orthorhombic, Tetragonal and Cubic phases of BaTiO$_3$. Around phase transitions, on a small temperature range $\approx$ 10 K, either one phase or the other can be obtained since all the points are obtained independently from each other (The MD is in each case initialized in the same way).}}
    \label{strain}
\end{figure}

The time averaged local mode components (averaged over the whole simulation cell) and the time averaged homogeneous strain tensor components are drawn on Figs.~\ref{polarization} and ~\ref{strain} as a function of temperature. In these two figures, we have renamed the components so that the polarization keeps the same direction for the whole range of temperature of a given phase (since all our points are obtained independently from each other, there is absolutely no reason that, for example, all the tetragonal configurations have their polarization along $z$).
As expected, we find the three phase transitions of BTO, well described as being first-order (both the polarization and the strain experience a discontinuity at the three phase transitions).

At low temperature, the three components of the macroscopic polarization are equal, as well as the three diagonal components of the homogeneous strain tensor. The non-diagonal components of the strain tensor are small and equal, but clearly non zero. This describes the rhombohedral (R) phase of barium titanate (space group $R3m$).

At about 190 K, one of the components of the macroscopic polarization falls to zero. In the same way, one of the diagonal components of the homogeneous strain tensor dissociates from the two others and becomes smaller. This describes the orthorhombic phase of barium titanate (space group $Amm2$). We note that the non-diagonal components of the strain tensor are not all equal to zero: $\eta_4$ (corresponding to $2 \epsilon_{yz}$) is not zero if $\eta_2$ ($\epsilon_{yy}$) and $\eta_3$ ($\epsilon_{zz}$) are the largest diagonal components.

At about 222-229 K, another component of the macroscopic polarization falls to zero. The homogeneous strain tensor has one diagonal component larger than the two others, the non-diagonal ones being zero. This is the tetragonal phase of barium titanate (space group $P4mm$).

Finally, at about 290-295 K (Curie temperature), all the components of the macroscopic polarization fall to zero, corresponding to the high-temperature cubic phase of barium titanate (space group $Pm \bar{3}m$). The diagonal components of the strain tensor are equal around 1.012. Since the strain is referenced in this hamiltonian to the LDA ground state of BTO (a$_0$ = 7.46 a.u.), this corresponds to a lattice constant of $\approx$ 3.995 {\AA}. All the non-diagonal components of the strain tensor are zero. We note that our simulation does not evidence any thermal expansion, which is one of the well-known problems of the Effective Hamiltonian\cite{tinte2003}.

The absolute values of polarization and homogeneous strains are very close to those of Zhong {\it et al}\cite{zhong94,zhong95}, while the phase transition temperatures are slightly lower ($\approx$ 290-295, 222-229 and 190 K in our case versus $\approx$ 300, 230 and 200 in the case of Zhong {\it et al}).

We now recalculate the temperature evolution of the polarization and strain of BaTiO$_3$, except that each temperature is now initialized with the result of a previous calculation. We perform these calculations by decreasing T from 360 to 160 K.

Practically, the two last configurations (including local and displacement modes and also strain tensor) are stored at the end of each calculation, so that the Verlet algorithm can be restarted with a new temperature. The calculations are still performed in the Nose-Hoover and Parinello-Rahman framework, with an external pressure -4.8 GPa, corresponding to a positive stress tensor $\sigma_{xx}=\sigma_{yy}=\sigma_{zz}=$4.8 GPa (and $\sigma_{xy}=\sigma_{xz}=\sigma_{yz}=0$). We perform 50000 steps for each temperature, with the same time step of 2$\times$10$^{-15}$s. The macroscopic averages are computed on the 40000 last steps.

\begin{figure}[htbp]
    {\par\centering
    {\scalebox{0.30}{\includegraphics{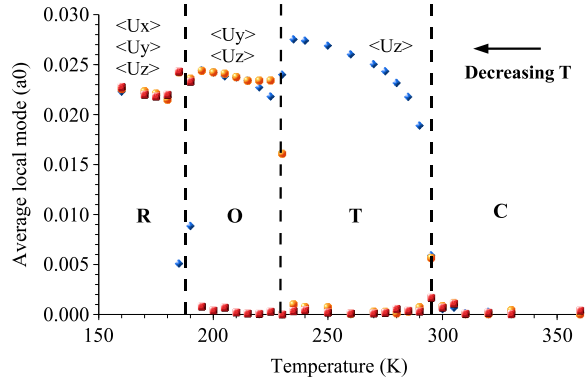}}}
    \par}
     \caption{{\small Average local mode components as a function of temperature. R, O, T and C denote respectively the Rhombohedral, Orthorhombic, Tetragonal and Cubic phases of BaTiO$_3$. Results obtained by decreasing the temperature and initializing the MD from last two configurations of the previous temperature. 50000 steps are performed for each point, and the averages computed on the last 40000.}}
    \label{polarization-descendant}
\end{figure}

\begin{figure}[htbp]
    {\par\centering
    {\scalebox{0.30}{\includegraphics{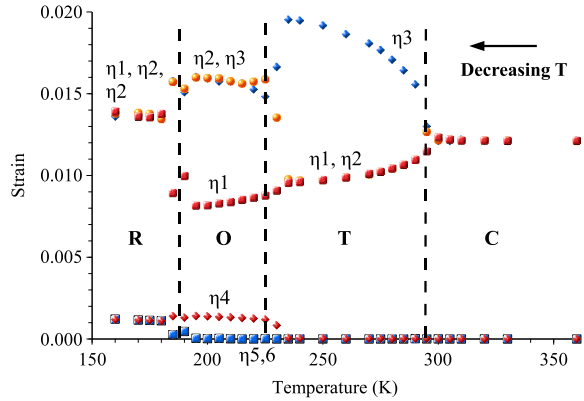}}}
    \par}
     \caption{{\small Average strain tensor components as a function of temperature (Voigt notation). R, O, T and C denote respectively the Rhombohedral, Orthorhombic, Tetragonal and Cubic phases of BaTiO$_3$. Results obtained by decreasing the temperature and initializing the MD from last two configurations of the previous temperature. 50000 steps are performed for each point, and the averages computed on the last 40000.}}
    \label{strain-descendant}
\end{figure}

As a consequence of this procedure, the temperature evolutions obtained are much more regular, especially close to phase transitions. Out of the regions close to phase transitions, the results are of course the same as in the previous case (within the numerical accuracy). The temperature evolution of polarization and strain are shown on Figs.~\ref{polarization-descendant} and ~\ref{strain-descendant} for the case of increasing temperature. The phase transitions are localized approximately at 190$\pm$5 (R-O), 230$\pm$5 (O-T) and 295$\pm$5 K (T-C).

\section{Conclusion and perspectives}

The MD method with Nose-Hoover and Parinello-Rahman schemes has been applied successfully to barium titanate in the framework of the Effective Hamiltonian\cite{zhong95}. A code has been implemented in fortran 90.
Thermal averages can be computed over the equilibrium trajectories obtained with these two techniques. This allows to reproduce the phase transitions and the temperature evolution of the polarization and strain tensor components of barium titanate, as it is found from Monte Carlo simulations\cite{zhong94,zhong95}. 
The code implemented can thus a priori be used with other Effective Hamiltonian parameters that can be found in the litterature. 

The MD simulations can be used to get insight, not only into thermal averages (strain, polarization, dielectric constant, etc) but also into dynamics and time-correlation, including fos instance dielectric loss\cite{ponomareva08}. Another very interesting possibility of MD is to perform thermodynamic integration to obtain free energy differences such as Landau free energies of ferroelectrics\cite{ti}.

\end{document}